\documentstyle[fleqn,11pt]{book}
\begin{document}
\parindent 1.4cm
\large
\begin{center}
{\Large \bf THE QUANTUM WAVE PACKET OF THE NON-LINEAR GROSS-PITAEVSKII
EQUATION}
\end{center}
\begin{center}
{{\bf J. M. F. Bassalo$^{1}$,\ P. T. S. Alencar$^{2}$,\  D. G. da
Silva$^{3}$,\ A. Nassar$^{4}$\ and\ M. Cattani$^{5}$}}
\end{center}
\begin{center}
{$^{1}$\ Funda\c{c}\~ao Minerva,\ Avenida Governador Jos\'e Malcher\ 629
- CEP\ 66035-100,\ Bel\'em,\ Par\'a,\ Brasil}
\end{center}
\begin{center}
{E-mail:\ jmfbassalo@gmail.com}
\end{center}
\begin{center}
{$^{2}$\ Universidade Federal do Par\'a\ -\ CEP\ 66075-900,\ Guam\'a,
Bel\'em,\ Par\'a,\ Brasil}
\end{center}
\begin{center}
{E-mail:\ tarso@ufpa.br}
\end{center}
\begin{center}
{$^{3}$\ Escola Munguba do Jari, Vit\'oria do Jari\ -\ CEP\
68924-000,\ Amap\'a,\ Brasil}
\end{center}
\begin{center}
{E-mail:\ danielgemaque@yahoo.com.br}
\end{center}
\begin{center}
{$^{4}$\ Extension Program-Department of Sciences, University of California,\
Los Angeles, California 90024,\ USA}
\end{center}
\begin{center}
{E-mail:\ nassar@ucla.edu}
\end{center}
\begin{center}
{$^{5}$\ Instituto de F\'{\i}sica da Universidade de S\~ao Paulo. C. P.
66318, CEP\ 05315-970,\ S\~ao Paulo,\ SP, Brasil}
\end{center}
\begin{center}
{E-mail:\ mcattani@if.usp.br}
\end{center}
\par
{\bf Abstract}:\ In this paper we study the quantum wave packet of the
non-linear Gross-Pitaeviskii equation.
\vspace{0.2cm}
\par
\vspace{0.2cm}
\par
{\bf 1.\ Introduction}
\vspace{0.2cm}
\par
In the present work we investigate the quantum wave packet of the
one-dimensional non-linear Gross-Pitaevskii equation with the potencial
$V(x,\ t)$ given by:
\begin{center}
{$V(x,\ t)\ =\ {\frac {1}{2}}\ m\ {\omega}^{2}(t)\ x^{2}$\ ,\ \ \ \ \ (1.1)}
\end{center}
which is the time dependent harmonic oscillator potential.
\par
2.\ {\bf Gross-Pitaeviskii Equation}
\par
Em 1961[1,2], E. P. Gross and, independently, L. P. Pitaevskii proposed a
non-linear Schr\"{o}dinger equation to represent time dependent
physical systems, given by:
\begin{center}
{i\ ${\hbar}\ {\frac {{\partial}{\psi}(x,\ t)}{{\partial}t}}\ =\
-\ {\frac {{\hbar}^{2}}{2\ m}}\ {\frac {{\partial}^{2}\ {\psi}(x,\
t)}{{\partial}x^{2}}}\ +\ {\frac {1}{2}}\ m\ {\omega}^{2}(t)\
x^{2}\ {\psi}(x,\ t)\ +\ g{\mid}\ {\psi}(x,\ t)\ {\mid}^{2}\
{\psi} (x,\ t)$\ ,\ \ \ \ \ (2.1)}
\end{center}
where ${\psi}(x,\ t)$ is a wavefunction and $g$ is a constant.
constante.
\par
Writting the wavefuncition ${\psi}(x,\ t)$ in the polar form, defined by
the Madelung-Bohm transformation[3,4], we get:
\begin{center}
{${\psi}(x,\ t)\ =\ {\phi}(x,\ t)\ e^{i\ S(x,\ t)}$\ ,\ \ \ \ \ (2.2)}
\end{center}
where $S(x\ ,t)$ is the classical action and ${\phi}(x,\ t)$ will be
defined in what follows.
\par
Substituting Eq.(2.2) into Eq.(2.1) and taking the real and imaginary
parts of the resulting equation, we get[5]:
\par
\begin{center}
{${\frac {{\partial}{\rho}}{{\partial}t}}\ +\ {\frac
{{\partial}({\rho}\ v_{qu})}{{\partial}x}}\ =\ 0$\ ,\ \ \ \ \ (2.3)}
\end{center}
\begin{center}
{${\hbar}\ {\frac {{\partial}S}{{\partial}t}}\ +\ {\frac {1}{2}}\ m\
v_{qu}^{2}(t)\ +\ {\frac {1}{2}}\ m\ {\omega}^{2}(t)\ x^{2}\ +\
V_{qu}\ +\ V_{GP}\ =\ 0$\ ,\ \ \ \ \ (2.4)}
\end{center}
\begin{center}
{${\frac {{\partial}v_{qu}}{{\partial}t}}\ +\ v_{qu}\ {\frac
{{\partial}v_{qu}}{{\partial}x}}\ +\ {\omega}^{2}(t)\ x\ =\ -\
{\frac {1}{m}}\ {\frac {{\partial}}{{\partial}x}}\ (V_{qu}\ +\ V_{GP})$\
,\ \ \ \ \ (2.5)}
\end{center}
where:
\begin{center}
{${\rho}(x,\ t)\ =\ {\phi}^{2}(x,\ t)$\ ,\ \ \ \ \ (2.6)\ \ \
(quantum mass density)}
\end{center}
\begin{center}
{$v_{qu}(x,\ t)\ =\ {\frac {{\hbar}}{m}}\ {\frac {{\partial}S(x,\
t)}{{\partial}x}}$\ ,\ \ \ \ \ (2.7)\ \ \ \ \ (quantum velocity)}
\end{center}
\begin{center}
{$V_{qu}(x,\ t)\ =\ -\ {\frac {{\hbar}^{2}}{2\ m}}\ {\frac {1}{{\sqrt
{{\rho}}}}}\ {\frac {{\partial}^{2}{\sqrt
{{\rho}}}}{{\partial}x^{2}}}\ =\ -\ {\frac {{\hbar}^{2}}{2\ m\
{\phi}}}\ {\frac {{\partial}^{2}{\phi}}{{\partial}x^{2}}}$\
,\ \ \ \ \ (2.8a,b)\ \ \ \ \ (Bohm quantum potential)}
\end{center}
and
\begin{center}
{$V_{GP}\ =\ g\ {\rho}$\ .\ \ \ \ \ (2.9)\ \ \ \ (Gross-Pitaevskii potential)}
\end{center}
\vspace{0.2cm}
\par
{\bf 3.\ Quantum Wave Packet}
\vspace{0.2cm}
\par
In 1909 [6], Einstein studied the black body radiation in thermodynamical
equilibrium with matter. Starting from Planck's equation, of 1900, of
the radiation density and using the Fourier expansion technique to
calculate its fluctuations, he showed that it exhibits, simultaneously,
fluctuations which are characteristic of waves and particles. In 1916
[7], analyzing again the black body Planckian radiation, Einstein
proposed that an electromagnetic radiation with wavelenght ${\lambda}$
had a linear momentum $p$, given by the relation:
\begin{center}
{p\ =\ ${\frac {h}{{\lambda}}}$\ ,\ \ \ \ \ (3.1)}
\end{center}
where {\bf h} is the Planck constant [8].
\par
In works developed between 1923 and 1925 [9] de Broglie formulated his
fundamental idea that the electron with mass $m$, in its atomic orbital
motion with velocity $v$ and linear momentum $p\ =\ m\ v$ is guided by
a "matter wave" (pilot-wave) with wavelenght is given by:
\begin{center}
{${\lambda}\ =\ {\frac {h}{p}}\ .\ \ \ \ \ (3.2)$}
\end{center}
\par
In 1926 [10], Schr\"{o}\-dinger proposed that the "pilot-wave de
Brogliean" ought to obey a differential equation, today know as the
famous Schr\"{o}dinger's equation:
\begin{center}
{$i\ {\hbar}\ {\frac {{\partial}}{{\partial}t}}\ {\Psi}({\vec {r}},\ t)
=\ {\hat {H}}\ {\Psi}({\vec {r}},\ t)$\ ,\ \ \ \ \ (3.3a)}
\end{center}
where ${\hat {H}}$ is Hamiltonian operator definied by:
\begin{center}
{${\hat {H}}\ =\ {\frac {{\hat {p}}^{2}}{2\ m}}\ +\ V({\vec {r}},\ t)\
,\ \ \ \ \ ({\hat {p}}\ =\ -\ i\ {\hbar}\ {\nabla})$\ ,\ \ \ \ \
(3.3b,c)}
\end{center}
where $V$ is the potential energy. In this same year of 1926 [11] Born
interpreted the Schr\"{o}\-dinger wave function ${\Psi}$ as being an
amplitude of probability.
\vspace{0.2cm}
\par
{\bf 4.\ The Quantum Wave Packet of the non-linear Gross-Pitaeviskii
equation}
\vspace{0.2cm}
\par
Initially, let us calculate the quantum trajectory ($x_{qu}$) of the
physical system represented by the eq.(2.1). To do this, let us
integrate the equation given by [12] [remember that q(t)\ =\
${\langle}\ x\ {\rangle}$]:
\begin{center}
{$v_{qu}(x,\ t)\ =\ {\frac {{\dot {{\sigma}}}(t)}{2\ {\sigma}(t)}}\ [x\
-\ q(t)]\ +\ {\dot {q}}(t)$\ ,\ \ \ \ \ (4.1)}
\end{center}
where:
\begin{center}
{${\sigma}^{2}(t)\ =\ {\frac {{\hbar}^{2}}{m^{2}}}\ {\frac {1}{k(t)\ +\
{\frac {2\ g}{{\sigma}(t)\ m\ {\sqrt {{\pi}\ {\sigma}(t)}}}}}}$\ ,\ \ \ \
\ (4.2)}
\end{center}
and:
\begin{center}
{$k(t)\ =\ [{\frac {{\hbar}^{2}}{m^{2}\ {\sigma}^{2}(t)}}\ -\ {\frac {2\
g}{{\sigma}(t)\ m\ {\sqrt {{\pi}\ {\sigma}(t)}}}}]$\ .\ \ \ \ \ (4.3)}
\end{center}
\par
So, remembering that ${\int}\ {\frac
{dz}{z}}\ =\ {\ell}n\ z,\ {\ell}n\ ({\frac {x}{y}})\ =\ {\ell}n\ x\ -\
{\ell}n\ y$\ , and \ ${\ell}n\ x\ y\ =\ {\ell}n\ x\ +\ {\ell}n\ y$], we have:
\begin{center}
{$v_{qu}(x,\ t)\ =\ {\frac {dx_{qu}}{dt}}\ =\ {\frac {{\dot
{{\sigma}}}(t)}{2\ {\sigma}(t)}}\ [x\ -\ q(t)]\ +\ {\dot {q}}(t)\ ,\ \
\ {\to}$}
\end{center}
\begin{center}
{${\frac {dx_{qu}}{dt}}\ -\ {\frac {dq}{dt}}\ =\ {\frac {{\dot
{{\sigma}}}(t)}{2\ {\sigma}(t)}}\ [x\ -\ q(t)]\ \ \ {\to}\ \ \ {\frac
{d[x_{qu}(t)\ -\ q(t)]}{[x_{qu}(t)\ -\ q(t)]}}\ =\ {\frac {{\dot
{{\sigma}}}(t)\ dt}{2\ {\sigma}(t)}}\ =\ {\frac {d{\sigma}(t)}{2\
{\sigma}(t)}}\ \ \ {\to}$}
\end{center}
\begin{center}
{${\int}_{o}^{t}\ {\frac {d[x_{qu}(t')\ -\ q(t')]}{[x_{qu}(t')\ -\
q(t')]}}\ =\ {\int}_{o}^{t}\ {\frac {d{\sigma}(t')}{2\ {\sigma}(t')}}\ \
\ {\to}$}
\end{center}
\begin{center}
{${\ell}n\ {\Big {(}}\ {\frac {[x_{qu}(t)\ -\ q(t)]}{[x_{qu}(0)\ -\
q(0)]}}\ {\Big {)}}\ =\ {\frac {1}{2}}\ {\ell}n\ {\Big {[}}\ {\frac
{{\sigma}(t)}{{\sigma}(0)}}\ {\Big {]}}\ =\ {\ell}n\ {\Big {[}}\ {\frac
{{\sigma}(t)}{{\sigma}(0)}}\ {\Big {]}}^{1/2}\ \ \ {\to}$}
\end{center}
\begin{center}
{$x_{qu}(t)\ =\ q(t)\ +\ {\Big {[}}\ {\frac {{\sigma}(t)}{{\sigma}(0)}}\
{\Big {]}}^{1/2}\ [x_{qu}(0)\ -\ q(0)]$\ ,\ \ \ \ \ (4.4)}
\end{center}
that represent the looked for quantum trajectory.
\par
To obtain the the quantum wave packet of the non-linear
Gross-Pitaeviskii equation given by the eq.(2.2), let us expand the
functions $S(x,\ t)$, $V(x,\ t)$ and $V_{qu}(x,\ t)$ around of $q(t)$
up to second Taylor order [5]. In this way we have:
\begin{center}
{$S(x,\ t)\ =\ S[q(t),\ t]\ +\ S'[q(t),\ t]\ [x\ -\ q(t)]\ +\ {\frac
{S''[q(t),\ t]}{2}}\ [x\ -\ q(t)]^{2}$\ ,\ \ \ \ \ (4.5)}
\end{center}
\begin{center}
{$V(x,\ t)\ =\ V[q(t),\ t]\ +\ V'[q(t),\ t]\ [x\ -\ q(t)]\ +\ {\frac
{V''[q(t),\ t]}{2}}\ [x\ -\ q(t)]^{2}$\ ,\ \ \ \ \ (4.6)}
\end{center}
\begin{center}
{$V_{qu}(x,\ t)\ =\ V_{qu}[q(t),\ t]\ +\ V_{qu}'[q(t),\ t]\ [x\ -\
q(t)]\ +\ {\frac {V_{qu}''[q(t),\ t]}{2}}\ [x\ -\ q(t)]^{2}$\ ,\ \ \ \
\ (4.7)}
\end{center}
\begin{center}
{$V_{GP}(x,\ t)\ =\ V_{GP}[q(t),\ t]\ +\ V_{GP}'[q(t),\ t]\ [x\ -\
q(t)]\ +\ {\frac {V_{GP}''[q(t),\ t]}{2}}\ [x\ -\ q(t)]^{2}$\ ,\ \ \ \
\ (4.8)}
\end{center}
where (') and ('') signifies, respectively, ${\frac
{{\partial}}{{\partial}q}}$\ and ${\frac
{{\partial}^{2}}{{\partial}q^{2}}}$.
\par
Differentiating the expression (4.5) in the variable $x$, multiplying
the result by ${\frac {{\hbar}}{m}}$, using the relations (2.6) and
(4.1), and taking into account the polynomial identity property, we
obtain:
\begin{center}
{${\frac {{\hbar}}{m}}\ {\frac {{\partial}S(x,\ t)}{{\partial}x}}\ =\
{\frac {{\hbar}}{m}}\ {\Big {(}}\ S'[q(t),\ t]\ +\ S''[q(t),\ t]\ [x\
-\ q(t)]\ {\Big {)}}\ =$}
\end{center}
\begin{center}
{$=\ v_{qu}(x,\ t)\ =\ {\big {[}}\ {\frac {{\dot
{{\sigma}}}(t)}{2\ {\sigma}(t)}}{\big {]}}\ [x\ -\ q(t)]\ +\ {\dot
{q}}(t)\ =\ \ \ {\to}$}
\end{center}
\begin{center}
{$S'[q(t),\ t]\ =\ {\frac {m\ {\dot {q}}(t)}{{\hbar}}}\ ,\ \ \
S''[q(t),\ t]\ =\ {\frac {m}{{\hbar}}}\ {\big {[}}\ {\frac {{\dot
{{\sigma}}}(t)}{2\ {\sigma}(t)}}\ {\big {]}}$\ .\ \ \ \ \ (4.9a,b)}
\end{center}
\par
Substituting the expressions (4.9a,b) in the equation (4.5),
results:
\begin{center}
{$S(x,\ t)\ =\ S_{o}(t)\ +\ {\frac {m\ {\dot {q}}(t)}{{\hbar}}}\ [x\ -\
q(t)]\ +\ {\frac {m}{4\ {\hbar}}}\ {\Big {[}}\ {\frac {{\dot
{{\sigma}}}(t)}{{\sigma}(t)}}\ {\Big {]}}\ [x\ -\ q(t)]^{2}$\ ,\ \ \ \
\ (4.10)}
\end{center}
where:
\begin{center}
{$S_{o}(t)\ {\equiv}\ S[q(t),\ t]$\ ,\ \ \ \ \ (4.11)}
\end{center}
is the quantum action.
\par
Differentiating the eq.(4.10) in relation to the time $t$, we obtain
(remembering that ${\frac {{\partial}x}{{\partial}t}}$\ =\ 0):
\begin{center}
{${\frac {{\partial}S}{{\partial}t}}\ =\ {\dot {S}}_{o}(t)\ +\ {\frac
{{\partial}}{{\partial}t}}\ {\Big {(}}\ {\frac {m\ {\dot
{q}}(t)}{{\hbar}}}\ [x\ -\ q(t)]\ {\Big {)}}\ +\ {\frac
{{\partial}}{{\partial}t}}\ {\Bigg {(}}\ {\frac {m}{4\ {\hbar}}}\ {\Big
{[}}\ {\frac {{\dot {{\sigma}}}(t)}{{\sigma}(t)}}\ {\Big {]}}\ [x\ -\
q(t)]^{2}\ {\Bigg {)}}\ \ \ {\to}$}
\end{center}
\begin{center}
{${\frac {{\partial}S}{{\partial}t}}\ =\ {\dot {S}}_{o}(t)\ +\ {\frac
{m\ {\ddot {q}}(t)}{{\hbar}}}\ [x\ -\ q(t)]\ -\ {\frac {m\ {\dot
{q}}(t)^{2}}{{\hbar}}}\ +$}
\end{center}
\begin{center}
{+\ ${\frac {m}{4\ {\hbar}}}\ [{\frac {{\ddot
{{\sigma}}}(t)}{{\sigma}(t)}}\ -\ {\frac {{\dot
{{\sigma}}}^{2}(t)}{{\sigma}^{2}(t)}}]\ [x\ -\ q(t)]^{2}\ -\ {\frac {m\
{\dot {q}}(t)}{2\ {\hbar}}}\ {\Big {(}}\ {\frac {{\dot
{{\sigma}}}(t)}{{\sigma}(t)}}\ {\Big {)}}\ [x\ -\ q(t)]$\ .\ \ \ \ \ (4.12)}
\end{center}
\par
Considering that [12]:
\begin{center}
{${\phi}(x,\ t)\ =\ {\sqrt {{\rho}(x,\ t)}}\ =\ [{\pi}\ {\sigma}(t)]^{-\
1/4}\ e^{-\ {\frac {[x\ -\ {\bar {x}}(t)]^{2}}{2\ {\sigma}(t)}}}\ \ \
{\to}$}
\end{center}
\begin{center}
{${\rho}(x,\ t)\ =\ [{\pi}\ {\sigma}(t)]^{-\
1/2}\ e^{-\ {\frac {[x\ -\ {\bar {x}}(t)]^{2}}{{\sigma}(t)}}}$,\ \ \ (4.13)}
\end{center}
let us write $V_{qu}$ in terms of $[x\ -\ q(t)]$. Initially, we
calculate the following differentiations:
\begin{center}
{${\frac {{\partial}{\phi}}{{\partial}x}}\ =\ {\frac
{{\partial}}{{\partial}x}}\ {\Big {(}}\ [{\pi}\ {\sigma}(t)]^{-\
1/4}\ e^{-\ {\frac {[x\ -\ q(t)]^{2}}{2\ {\sigma}(t)}}}\ {\Big
{)}}\ =\ [{\pi}\ {\sigma}(t)]^{-\ 1/4}\ e^{-\ {\frac {[x\ -\
q(t)]^{2}}{2\ {\sigma}(t)}}} {\frac {{\partial}}{{\partial}x}}\
{\Big {(}}\ -\ {\frac {[x\ -\ q(t)]^{2}}{2\ {\sigma}(t)}}\ {\Big
{)}}\ \ \ {\to}$}
\end{center}
\begin{center}
{${\frac {{\partial}{\phi}}{{\partial}x}}\ =\ -\ [{\pi}\
{\sigma}(t)]^{-\ 1/4}\ e^{-\ {\frac {[x\ -\ q(t)]^{2}}{2\
{\sigma}(t)}}}\ {\frac {[x\ -\ q(t)]}{{\sigma}(t)}}$\ ,}
\end{center}
\begin{center}
{${\frac {{\partial}^{2}{\phi}}{{\partial}x^{2}}}\ =\ {\frac
{{\partial}}{{\partial}x}}\ {\Big {(}}\ -\ [{\pi}\ {\sigma}(t)]^{-\
1/4}\ e^{-\ {\frac {[x\ -\ q(t)]^{2}}{2\ {\sigma}(t)}}}\ {\frac
{[x\ -\ q(t)]}{{\sigma}(t)}}\ {\Big {)}}$\ =}
\end{center}
\begin{center}
{$\ =\ -\ [{\pi}\ {\sigma}(t)]^{-\ 1/4}\ e^{-\ {\frac {[x\ -\
q(t)]^{2}}{2\ {\sigma}(t)}}}\ {\frac {{\partial}}{{\partial}x}}\
{\Big {(}}\ {\frac {[x\ -\ q(t)]}{{\sigma}(t)}}\ {\Big {)}}\ -$}
\end{center}
\begin{center}
{$-\ [{\pi}\ {\sigma}(t)]^{-\ 1/4}\ e^{-\ {\frac {[x\ -\
q(t)]^{2}}{2\ {\sigma}(t)}}}\ {\frac {{\partial}}{{\partial}x}}\
{\Big {(}}\ {\frac {[x\ -\ q(t)]^{2}}{2\ {\sigma}(t)}}\ {\Big
{)}}\ {\big {(}}\ {\frac {[x\ -\ q(t)]}{{\sigma}(t)}}\ {\big
{)}}\ \ \ {\to}$}
\end{center}
\begin{center}
{${\frac {{\partial}^{2}{\phi}}{{\partial}x^{2}}}\ =\ -\ [{\pi}\
{\sigma}(t)]^{-\ 1/4}\ e^{-\ {\frac {[x\ -\ q(t)]^{2}}{2\
{\sigma}(t)}}}\ {\frac {1}{{\sigma}(t)}}\ +\ [{\pi}\
{\sigma}(t)]^{-\ 1/4}\ e^{-\ {\frac {[x\ -\ q(t)]^{2}}{2\
{\sigma}(t)}}}\ {\frac {[x\ -\ q(t)]^{2}}{2\ {\sigma}^{2}(t)}}$\ =}
\end{center}
\begin{center}
{$=\ -\ {\phi}\ {\frac {1}{{\sigma}(t)}}\ +\ {\phi}\ {\frac {[x\
-\ q(t)]^{2}}{{\sigma}^{2}(t)}}\ \ \ {\to}\ \ \ {\frac {1}{{\phi}}}\
{\frac {{\partial}^{2}{\phi}}{{\partial}x^{2}}}\ =\ -\ {\frac {2\
m}{{\hbar}^{2}}}\ V_{qu}(x,\ t)\ =\  {\frac {[x\ -\
q(t)]^{2}}{{\sigma}^{2}(t)}}\ -\ {\frac {1}{{\sigma}(t)}}$\ \ \ \ \ \
(4.14)}
\end{center}
\par
Substituting the relation (4.14) in the equation (2.7b), taking into
account the expression (4.7), results:
\begin{center}
{$V_{qu}(x,\ t)\ =\ V_{qu}[q(t),\ t]\ +\ V_{qu}'[q(t),\ t]\ [x\ -\
q(t)]\ +\ {\frac {V_{qu}''[q(t),\ t]}{2}}\ [x\ -\ q(t)]^{2}\ \ \ {\to}$}
\end{center}
\begin{center}
{$V_{qu}(x,\ t)\ =\ {\frac {{\hbar}^{2}}{2\ m\ {\sigma}(t)}}\ [x\
-\ q(t)]^{o}\ -\ {\frac {{\hbar}^{2}}{2\ m\ {\sigma}^{2}(t)}}\ [x\ -\
q(t)]^{2}$\ \ \ \ \ (4.15)}
\end{center}
\par
Comparing the eqs. (4.7) and (4.15), we have:
\begin{center}
{$V_{qu}[q(t),\ t]\ =\ {\frac {{\hbar}^{2}}{2\ m\ {\sigma}(t)}};\ \ \
V_{qu}'[q(t),\ t]\ =\ 0;\ \ \ V_{qu}"[q(t),\ t]\ =\ -\  {\frac
{{\hbar}^{2}}{m\ {\sigma}^{2}(t)}}.$\ \ \ (4.16a,b,c)}
\end{center}
\par
Besides this the eq.(4.6) will be written, using the eq.(1.1) in the
form:
\begin{center}
{$V(x,\ t)\ =\ V[q(t),\ t]\ +\ V'[q(t),\ t]\ [x\ -\ q(t)]\ +\ {\frac
{V''[q(t),\ t]}{2}}\ [x\ -\ q(t)]^{2}\ \ \ {\to}$}
\end{center}
\begin{center}
{$V(x,\ t)\ =\ {\frac {1}{2}}\ m\ {\omega}^{2}(t)\ q^{2}(t)\ +$}
\end{center}
\begin{center}
{$+\ {\Big {(}}\ m\ {\omega}^{2}(t)\ q(t)\ {\Big {)}}\ [x\ -\ q(t)]\ +\
{\frac {m}{2}}\ {\omega}^{2}(t)\ [x\ -\ q(t)]^{2}$\ .\ \ \ \ \
(4.17)}
\end{center}
\par
Comparing the eqs. (4.6) and (4.17), results:
\begin{center}
{$V[q(t),\ t]\ =\ {\frac {1}{2}}\ m\ {\omega}^{2}(t)\ q^{2}(t);\ \ \
V'[q(t),\ t]\ =\ m\ {\omega}^{2}(t)\ q(t)$;\ \ \ (4.18a,b)}
\end{center}
\begin{center}
{$V"[q(t),\ t]\ =\ m\ {\omega}^{2}(t).$\ \ \ (4.18c)}
\end{center}
\par
Now,\ let us expand the eq. (4.13) around of q(t) to second Taylor
order:
\begin{center}
{${\rho}(x,\ t)\ =\ {\rho}[q(t),\ t]\ +\ {\rho}'[q(t),\ t]\ [x\ -\
q(t)]\ +\ {\frac {{\rho}''[q(t),\ t]}{2}}\ [x\ -\ q(t)]^{2}\ \ \ {\to}$}
\end{center}
\begin{center}
{${\rho}(x,\ t)\ =\ [{\pi}\ {\sigma}(t)]^{-\
1/2}\ e^{-\ {\frac {[x\ -\ q(t)]^{2}}{{\sigma}(t)}}}\ +\
[{\pi}\ {\sigma}(t)]^{-\ 1/2}\ e^{-\ {\frac {[x\ -\
q(t)]^{2}}{{\sigma}(t)}}}\ {\frac {2\ [x\ -\ q(t)]^{2}}{{\sigma}(t)}}$\
+}
\end{center}
\begin{center}
{$+\ [{\pi}\ {\sigma}(t)]^{-\ 1/2}\ e^{-\ {\frac {[x\ -\
q(t)]^{2}}{{\sigma}(t)}}}\ {\Big {(}}\ -\ {\frac {1}{{\sigma}(t)}}\ +\
{\frac {2\ [x\ -\ q(t)]^{2}}{{\sigma}^{2}(t)}}\ {\Big {)}}\
[x\ -\ q(t)]^{2}$.\ }
\end{center}
\par
Considering only quadratics terms, results:
\begin{center}
{${\rho}(x,\ t)\ =\ [{\pi}\ {\sigma}(t)]^{-\
1/2}\ e^{-\ {\frac {[x\ -\ q(t)]^{2}}{{\sigma}(t)}}}\ {\Big {(}}\ 1\
+\ {\frac {[x\ -\ q(t)]^{2}}{{\sigma}(t)}}\ {\Big {)}}$.\ \ \ \ \ (4.19)}
\end{center}
\par
By using the eqs. (2.9), (4.8) and (4.19), we have:
\begin{center}
{$V_{GP}[q(t),\ t]\ =\ g\ [{\pi}\ {\sigma}(t)]^{-\
1/2}\ e^{-\ {\frac {[x\ -\ q(t)]^{2}}{{\sigma}(t)}}}\ {\Big {(}}\ 1\
+\ {\frac {[x\ -\ q(t)]^{2}}{{\sigma}(t)}}\ {\Big {)}}$\ =}
\end{center}
\begin{center}
{$=\ V_{GP}[q(t),\ t]\ +\ V_{GP}'[q(t),\ t]\ [x\ -\
q(t)]\ +\ {\frac {V_{GP}''[q(t),\ t]}{2}}\ [x\ -\ q(t)]^{2}\ \ \ {\to}$}
\end{center}
\begin{center}
{$V_{GP}[q(t),\ t]\ =\ g\ [{\pi}\ {\sigma}(t)]^{-\
1/2}\ e^{-\ {\frac {[x\ -\ q(t)]^{2}}{{\sigma}(t)}}};\ \ \
V'_{GP}[q(t),\ t]\ =\ 0$;\ \ \ \ \ (4.20a,b)}
\end{center}
\begin{center}
{$V"_{GP}[q(t),\ t]\ =\ {\frac {2\ g\ [{\pi}\ {\sigma}(t)]^{-\
1/2}\ e^{-\ {\frac {[x\ -\ q(t)]^{2}}{{\sigma}(t)}}}}{{\sigma(t)}}}$. \
\ \ \ (4.20c)}
\end{center}
\par
Inserting the eqs. (2.7), (4.1;\ 4.5-8; 4.12;\ 4.14;\ 4.16a-c;\
4.18a-c;\ 4.20a-c) into the eq.(2.4), we obtain, remembering that
$S_{o}(t)$, ${\sigma}(t)$ and $q(t)$:
\begin{center}
{${\hbar}\ {\frac {{\partial}S}{{\partial}t}}\ +\ {\frac {1}{2}}\ m\
v_{qu}^{2}\ +\ V\ +\ V_{qu}\ +\ V_{GP}\ =$}
\end{center}
\begin{center}
{$=\ {\hbar}\ {\Big {[}}\ {\dot {S}}_{o}\ +\ {\frac {m\ {\ddot
{q}}}{{\hbar}}}\ (x\ -\ q)\ -\ {\frac {m\ {\dot
{q}}^{2}}{{\hbar}}}\ +\ {\frac {m}{4\ {\hbar}}}\ {\big {(}}\ {\frac
{{\ddot {{\sigma}}}}{{\sigma}}}\ -\ {\frac {{\dot
{{\sigma}}}^{2}}{{\sigma}^{2}}}\ {\big {)}}\ (x\ -\ q)^{2}\ -$}
\end{center}
\begin{center}
{$-\ {\frac {m\ {\dot {q}}}{2\ {\hbar}}}\ {\big {(}}\ {\frac {{\dot
{{\sigma}}}}{{\sigma}}}\ {\big {)}}\ (x\ -\
q) {\Big {]}}\ +\ {\frac {1}{2}}\ m\ {\Big {[}}\ {\big {(}}\ {\frac
{{\dot {{\sigma}}}}{2\ {\sigma}}}\ {\big {)}}\ (x\ -\ q)\ +\ {\dot
{q}}\ {\Big {]}}^{2}\ +$}
\end{center}
\begin{center}
{$+\ {\frac {1}{2}}\ m\ {\omega}^{2}\ q^{2}\ +\ m\ {\omega}^{2}\
q^{2}\ (x\ -\ q)\ +\ {\frac {m}{2}}\ {\omega}^{2}\ (x\ -\ q)^{2}\ +\
{\frac {{\hbar}}{2\ m\ {\sigma}}}\ -$}
\end{center}
\begin{center}
{$-\ {\frac {{\hbar}^{2}}{2\ m\ {\sigma}^{2}}}\ (x\ -\ q)^{2}\ +\ g\
[{\pi}\ {\sigma}(t)]^{-\ 1/2}\ e^{-\ {\frac {[x\ -\
q]^{2}}{{\sigma}}}}\ {\Big {(}}\ 1\ +\ {\frac {2\ [x\ -\
q]^{2}}{{\sigma}}}\ {\Big {)}}=\ 0$\ .\ \ \
\ \ (4.21)}
\end{center}
\par
Since $(x\ -\ q)^{o}\ =\ 1$, we can gather together the above
expression in potencies of $(x\ -\ q)$, obtaining:
\begin{center}
{${\Big {[}}\ {\hbar}\ {\dot {S}}_{o}\ -\ m\ {\dot {q}}^{2}\ +\
{\frac {1}{2}}\ m\ {\dot {q}}^{2}\ +\ {\frac {1}{2}}\ m\
{\omega}^{2}\ q^{2} +\ {\frac {{\hbar}^{2}}{2\ m\ {\sigma}}}\ +\
g\ [{\pi}\ {\sigma}]^{-\ 1/2}\ e^{-\ {\frac {[x\ -\
q]^{2}}{{\sigma}}}}\ {\Big {]}}\ (x\ -\ q)^{o}\ +$}
\end{center}
\begin{center}
{$+\ {\Big {[}}\ m\ {\ddot {q}}\ -\ m\ {\dot {q}}\ {\frac {{\dot
{{\sigma}}}}{2\ {\sigma}}}\ +\  m\ {\dot {q}}\ {\frac {{\dot
{{\sigma}}}}{2\ {\sigma}}}\ +\ m\ {\omega}^{2}\ q\ {\Big {]}}\ (x\ -\
q)\ +\ {\Big {[}}\ {\frac {m\ {\ddot {{\sigma}}}}{4\ {\sigma}}}\ -\
{\frac {m\ {\dot {{\sigma}}}^{2}}{4\ {\sigma}^{2}}}\ +$}
\end{center}
\begin{center}
{$+\ {\frac {m\ {\dot {{\sigma}}}^{2}}{8\ {\sigma}^{2}}}\ +\ {\frac
{m\ {\omega}^{2}}{2}}\ -\ {\frac {{\hbar}^{2}}{2\ m\ {\sigma}^{2}}}\ +\
{\frac {2\ g\ [{\pi}\ {\sigma}(t)]^{-\ 1/2}\ e^{-\ {\frac {[x\ -\
q(t)]^{2}}{{\sigma}(t)}}}}{{\sigma(t)}}} {\Big {]}}\ (x\ -\ q)^{2}\ =\
0$\ .\ \ \ \ \ (4.22)}
\end{center}
\par
As the above relation is an identically null polynomium, the
coefficients of the potencies must be all equal to zero, that is:
\begin{center}
{${\dot {S}}_{o}(t)\ =\ {\frac {1}{{\hbar}}}\ {\Big {[}}\ {\frac
{1}{2}}\ m\ {\dot {q}}^{2}\ -\ {\frac {1}{2}}\ m\ {\omega}^{2}\
q^{2}\ -\ {\frac {{\hbar}^{2}}{2\ m\ {\sigma}^{2}}}\ -\ g\ [{\pi}\
{\sigma}]^{-\ 1/2}\ e^{-\ {\frac {[x\ -\ q]^{2}}{{\sigma}}}}\ {\Big
{]}}\ \ \ {\to}$}
\end{center}
\begin{center}
{${\dot {S}}_{o}(t)\ =\ {\frac {1}{{\hbar}}}\ {\Big {[}}\ {\frac
{1}{2}}\ m\ {\dot {q}}^{2}\ -\ {\frac {1}{2}}\ m\ {\omega}^{2}\
q^{2}\ -\ {\frac {{\hbar}^{2}}{2\ m\ {\sigma}^{2}}}\ -\ g\ {\rho}\ {\Big
{]}}$, \ \ \ \ \ (4.23)}
\end{center}
\begin{center}
{${\ddot {q}}\ +\ {\omega}^{2}(t)\ q\ =\ 0$\ ,\ \ \ \ \ (4.24)}
\end{center}
\begin{center}
{${\frac {{\ddot {{\sigma}}}}{2\ {\sigma}}}\ -\ {\frac {{\dot
{{\sigma}}^{2}}}{4\ {\sigma}^{2}}}\ +\ {\omega}^{2}\ +\ {\frac {2\ g\
{\rho}}{m\ {\sigma}}}\ =\ {\frac {{\hbar}^{2}}{m^{2}\ {\sigma}^{2}}}$\
.\ \ \ \ \ (4.25)}
\end{center}
\par
Assuming that the following initial conditions are obeyed:
\begin{center}
{$q(0)\ =\ x_{o}\ ,\ \ \ {\dot {q}}(0)\ =\ v_{o}\ ,\ \ \ {\sigma}(0)\
=\ a_{o}\ ,\ \ \ {\dot {{\sigma}}}(0)\ =\ b_{o}$\ ,\ \ \ \ \ \
(4.26a-d)}
\end{center}
and that [see eq.(2.7)]:
\begin{center}
{$S_{o}(0)\ =\ {\frac {m\ v_{o}\ x_{o}}{{\hbar}}}$\ ,\ \ \ \ \ (4.27)}
\end{center}
the integration of the expression (4.23) will be given by:
\begin{center}
{$S_{o}(t)\ =\ {\frac {1}{{\hbar}}}\ {\int}_{o}^{t}\ dt'\ {\Big {[}}\
{\frac {1}{2}}\ m\ {\dot {q}}^{2}(t')\ -\ {\frac {1}{2}}\ m\
{\omega}^{2}(t')\ q^{2}(t') -$}
\end{center}
\begin{center}
{$-\ {\frac {{\hbar}^{2}}{2\ m\ {\sigma}(t')}}\ -\ g\ {\rho}(x,\
t')\ {\Big {]}}\ +\ {\frac {m\ v_{o}\ x_{o}}{{\hbar}}}$\
.\ \ \ \ \ (4.28)}
\end{center}
\par
Taking into account the expressions (4.9a,b) and (4.28) in the
equation (4.10) results:
\begin{center}
{$S(x,\ t)\ =\ {\frac {1}{{\hbar}}}\ {\int}_{o}^{t}\ dt'\ {\Big {[}}\
{\frac {1}{2}}\ m\ {\dot {q}}^{2}(t')\ -\ {\frac {1}{2}}\ m\
{\omega}^{2}(t')\ q^{2}(t')\ -\ {\frac {{\hbar}^{2}}{2\ m\
{\sigma}(t')}}\ -\ g\ {\rho}(x,\ t')\  {\Big {]}}\ +$}
\end{center}
\begin{center}
{$+\ {\frac {m\ v_{o}\ x_{o}}{{\hbar}}}\ +\ {\frac {m\ {\dot
{q}}(t)}{{\hbar}}}\ [x\ -\ q(t)]\ +\ {\frac {m\ {\dot {\sigma}}}{4\
{\hbar}\ {\sigma}}}\ [x\ -\ q(t)]^{2}$\ .\ \ \ \ \ (4.29)}
\end{center}
\par
This result obtained above permit us, finally, to obtain the wave
packet for of the non-linear Gross-Pitaeviskii equation. Indeed,
considering the equations (2.2), (.6), (4.13) and (4.29), we get:
\begin{center}
{${\Psi}(x,\ t)\ =\ [{\pi}\ {\sigma}(t)]^{-\ 1/4}\ exp\ {\Bigg
{[}}\ {\Big {(}}\ {\frac {i\ m\ {\dot {{\sigma}}}(t)}{4\ {\hbar}\
{\sigma}(t)}}\ -\ {\frac {1}{2\ {\sigma}(t)}}\ {\Big {)}}\ [x\ -\
q(t)]^{2}\ {\Bigg {]}}\ {\times}$}
\end{center}
\begin{center}
{${\times}\ exp\ {\Big {[}}\ {\frac {i\ m\ {\dot {q}}(t)}{{\hbar}}}\ [x\
-\ q(t)]\ +\ {\frac {i\ m\ v_{o}\ x_{o}}{{\hbar}}}\ {\Big {]}}\ {\times}$}
\end{center}
\begin{center}
{${\times}\ exp\ {\Bigg {[}}\ {\frac {i}{{\hbar}}}\
{\int}_{o}^{t}\ dt'\ {\Big {[}}\ {\frac {1}{2}}\ m\ {\dot
{q}}^{2}(t')\ -\ {\frac {1}{2}}\ m\ {\omega}^{2}(t')\ q^{2}(t')
-\ {\frac {{\hbar}^{2}}{2\ m\ {\sigma}(t')}}\ -\ g\ {\rho}(x,\ t')
{\Big {]}}\ {\Bigg {]}}$\ .\ \ \ \ \ (4.30)}
\end{center}
\par
The eq. (4.30) we show that when $q\ =\ 0$, then we obtains the eq.
(3.3.2.25) of the Reference [5], if ${\sigma}(t)\ =\ 2\ a^{2}(t)$,\
$q(t)\ =\ X(t)$\ and ${\frac {1}{2}}\ m\ {\omega}^{2}(t)\ q^{2}(t)\ =\
V[X(t)]$.
\par
\begin{center}
\newpage
 {{\bf NOTES AND REFERENCES}}
\end{center}
\par
1.\ GROSS, E. P. 1961. {\it Nuovo Cimento 20}, 1766.
\par
2.\ PITAEVSKII, L. P. 1961. {\it Soviet Physics (JETP) 13}, 451.
\par
3.\ MADELUNG, E. 1926. {\it Zeitschrift f\"{u}r Physik 40}, 322.
\par
4.\ BOHM, D. 1952. {\it Physical Review 85}, 166.
\par
5.\ BASSALO, J. M. F., ALENCAR, P. T. S., CATTANI, M. S. D. e
NASSAR, A. B. 2003. {\bf T\'opicos da Mec\^anica Qu\^antica de de
Broglie-Bohm}, EDUFPA.
\par
6.\ EINSTEIN, A. 1909. {\it Physikalische Zeitschrift} {\bf 10}, p. 185.
\par
7.\ EINSTEIN, A. 1916. {\it Verhandlungen der Deutschen Physikalischen
Gesellschaft} {\bf 18}, p. 318; ----- 1916. {\it Mitteilungen der
Physikalischen Gesellschaft zu Z\"{u}rich} {\bf 16}, p. 47.
\par
8.\ This dual character of the eletromagntic radiation has been
just proposed by Stark, in 1909, in the paper published in the
{\it Physikalische Zetischrift} {\bf 10}, p. 902. In this paper
he explained bremsstrahlung.
\par
9.\ DE BROGLIE, L. 1923. {\it Comptes Rendus de l'Academie des
Sciences de Paris} {\bf 177}, pgs. 507; 548; 630; ----- 1924. {\it
Comptes Rendus de l'Academie des Sciences de Paris} {\bf 179}, p. 39;
----- 1925. {\it Annales de Physique} {\bf 3}, p. 22.
\par
10.\ SCHR\"{O}DINGER, E. 1926. {\it Annales de Physique Leipzig} {\bf
79}, pgs. 361; 489; 734; 747.
\par
11.\ BORN, M. 1926. {\it Zeitschrift f\"{u}r Physik} {\bf 37; 38}, pgs.
863; 803.
\par
12.\ BASSALO, J. M. F., ALENCAR, P. T. S., SILVA, D. G., NASSAR, A. B.
and CATTANI, M. {\it arXiv:0902.3125}\ [math-ph]\ 18\ February\ 2009.
\end{document}